\documentclass[11pt]{article}
\usepackage{times} 
\usepackage{mathptmx} 
\usepackage[utf8]{inputenc}
\usepackage{graphicx}
\usepackage{amsmath}
\usepackage[a4paper, total={6.5in, 10in}]{geometry}
\usepackage[T1]{fontenc}
\usepackage{bm}
\usepackage{xcolor}
\usepackage{abstract}

\newcommand \keywords[1]{\textbf{Keywords}:#1} 

\begin{document}

\title{\vspace{-2cm}Unconventional superconductivity in magic-strain graphene superlattices} 

\author{Qingxiang Ji$^{1 \dagger}$, Bohan Li$^{1\dagger}$, Johan Christensen$^2$, Changguo Wang$^{1*}$, Muamer Kadic$^{3*}$}
\date{$^1$National Key Laboratory of Science and Technology on Advanced Composites in Special Environments, Harbin Institute of Technology, Harbin, 150001, PR China\\
$^2$ IMDEA Materials Institute, Calle Eric Kandel 2
28906, Getafe (Madrid), Spain \\
$^3$Université de Franche-Comté, Institut FEMTO-ST, CNRS, 25000 Besançon, France\\
$\dagger$ These authors contributed equally to this work.\\
$^{*}$Corresponding authors.
E-mail: wangcg@hit.edu.cn; muamer.kadic@femto-st.fr\\}

\maketitle
\begin{onecolabstract}
  \vspace{-1cm} \normalsize \noindent Extensive investigations on the Moiré magic-angle have been conducted in twisted bilayer graphene, unlocking the mystery of unconventional superconductivity and insulating states. In analog to magic angle, here we demonstrate the new concept of magic-strain in graphene systems by judiciously tailoring mechanical relaxation (stretch and compression) which is easier to implement in practice. We elucidate the interplay of strain-induced effects and delve into the resulting unconventional superconductivity or semimetal-insulator transition in relaxation-strained graphene, going beyond the traditional twisting approach. Our findings reveal how relaxation strain can trigger superconducting transitions (with an ultra-flat band at the Fermi level) or the semimetal-insulator transition (with a gap opening at the $K$ point of $0.39\rm{~eV}$) in both monolayer and bilayer graphene. These discoveries open up a new branch for correlated phenomena and provide deeper insights into the underlying physics of superconductors, which positions graphene as a highly tunable platform for novel electronic applications.
\end{onecolabstract}
\thispagestyle{empty}

\section*{Introduction}
\noindent Functional materials are engineered materials designed with specific functionalities in mind. They play a crucial role in various technological advancements to harness sound \cite{christensen2007collimation,fleury2015invisible,cummer2016controlling,yu2018magnetoactive,jimenez2016ultra}, light \cite{schittny2014invisibility,farhat2013exciting}, vibrations \cite{wang2016tunable,chen2021realization}, heat \cite{li2021transforming,schittny2013experiments} and electronic states \cite{gansel2009gold,amin2021thz}. For many years, classical metamaterials have been the workhorse of the functional materials field. By manipulating their artificial structure features at the subwavelength scale, metamaterials achieve an array of exotic properties not found in natural materials \cite{wegener2013metamaterials,kadic20193d,bertoldi2017flexible}. For instance, metamaterials can bend light or sound in unusual ways, create invisibility cloaks, or possess negative refractive index \cite{smith2004metamaterials,schurig2006metamaterial,zhu2014negative,farhat2009ultrabroadband}. 

In recent years, a new class of functional materials has emerged – van der Waals (vdW) metamaterials. As the name suggests, vdW metamaterials are the marriage of vdW materials and metamaterial design principles. They create intricate heterostructures by stacking different vdW materials. These heterostructures can be tailored to exhibit entirely new properties due to the combined effects of their individual components \cite{geim2013van,novoselov20162d,meng2023photonic}, opening doors to novel functionalities not achievable by classical metamaterials alone. vdW metamaterials have shown great potential as tunable correlated electron systems, and have demonstrated various intriguing properties by varying the stacking configuration of low-dimension material sheets, e.g., graphene and Mexenes \cite{zhao2023ultrathin,qin2023graphene,christensen2012graphene,ponomarenko2013cloning,san2012non}. For graphene, the emergent heterostructures have added them a long list of miraculous properties such as the superconducting and insulating state.
Compared to other superconducting materials with intense doping \cite{ichinokura2016superconducting}, e.g., copper oxide \cite{massee2019noisy}, iron-based \cite{liu2023pair} and MgB$_2$ superconductors \cite{jin2019topological}, graphene has unique advantages of being a single-atomic lattice structure. This superlattice characteristic expands possibilities to tune graphene's conductivity properties by tailoring its heterostructure using mechanical deformation/strain. 
Strain is an effective way for engineering flat bands that favor the emergence of superconductivity or other correlated phases \cite{guinea2010energy,klimov2012electromechanical,lu2012transforming,zabel2012raman,amorim2016novel,mao2020evidence,li2020efficient,yang2021strain, mahmud2023topological}.
Recently experiments have demonstrated  superconducting states in twisted bilayer graphene (TBG) \cite{cao2018unconventional}.  Superconductivity of twisted graphene systems is rooted in Moiré-modulation of the interlayer coupling, which is depicted by Dirac models that flatten the electronic bands at particular angles \cite{bistritzer2011moire,tarnopolsky2019origin}.
The fascinating physics of correlated graphene Moiré superlattices, such as TBG, has generated extensive efforts to uncover the mysteries of their phase diagrams \cite{chen2023anomalous}. As a typical example, the independent-layer behavior and the reduction of the Fermi velocity are observed for small angles in TBG. Specifically, when the torsional angle is close to 1.1° (magic angle), superconductivity and Mott insulator behavior can be induced in TBG. In addition, magic angles can also cause some exotic phenomena in optics and mechanics \cite{hu2020topological,gonzalez2017electrically}. So far, an outburst of research has been conducted on twisting modulation. However, accurate twisting is laborious and needs intense efforts during sample fabrication. The influence of in-plane stain/deformation generated during the twisting process is also neglected   \cite{park2021tunable,ji2023interlayer,morovati2022interlayer}.

Compared with twisting, relaxation (stretch or compression), which is widely adopted in mechanics, is easier to implement and holds potential for large-scale device applications.
Researchers demonstrate that modulating relaxation strain can generate an approximate flat-band state or induce a bandgap in monolayer graphene, similar to those produced in twisted bilayer configurations. 
Experiments that engineer relaxation strain on graphene membranes have reported unexpected electronic transport and peculiar local density of states features \cite{banerjee2020strain,jessen2019lithographic}. Although intriguing phenomena have been predicted, there is a gap in connecting the unconventional properties to distinct strain behavior. Knowledge of the strain features that determine the resulting electronic properties is highly desirable. Currently various strain conditions have been implemented in monolayer graphene, but the bandgap tunability is relatively confined. For bilayer graphene, the strain effects are studied all within the framework of twisted conditions \cite{guinea2018electrostatic,nakatsuji2022moire,zhang2023effects}. We note that the using bi-axial relaxation strain on graphene systems (monolayer and bilayer) to achieve unconventional properties, which is predicted to have a higher degree of tunability (meanwhile more complicated), remains elusive. Consequently, the potential of relaxation-strained graphene to tailor electronic properties remains untapped.

In this work, we address these issues by developing tight-binding models that control bi-axial strain on graphene sheets. Firstly, we study monolayer graphenes with symmetrical strain distribution and demonstrate that relaxation will influence the Fermi velocity near the $K$ point. Based on this finding, we adopt a general deformation manner where the graphene is stretched in one direction and compressed in the perpendicular direction. This technique allows us to open the bandgap largely ($0.39\rm{~eV}$) and generate a semimetal-insulator transition for monolayer graphene. Then we turn to investigate Bernal-stacked graphene and reveal the relationship between interlayer distance and false Van der Walls force for bilayer graphene systems. By fixing one graphene layer and stretching another layer with a symmetrical strain rate of 1.9$\%$ (magic strain), we display unambiguously a flat band at the Fermi level, indicating a superconducting transition. In addition, the asymmetric strain on bilayer graphene will open the bandgap with small margins ($0.0272\rm{~eV}$), much less than the monolayer counterpart ($0.39\rm{~eV}$).

\section*{Material and methods}
\noindent We construct a tight-binding model (TBM) for monolayer graphene sheets, based on which we investigate the band structure considering two types of strain distributions:\\

\begin{figure}[!ht]
\centering
\includegraphics[width=\linewidth]{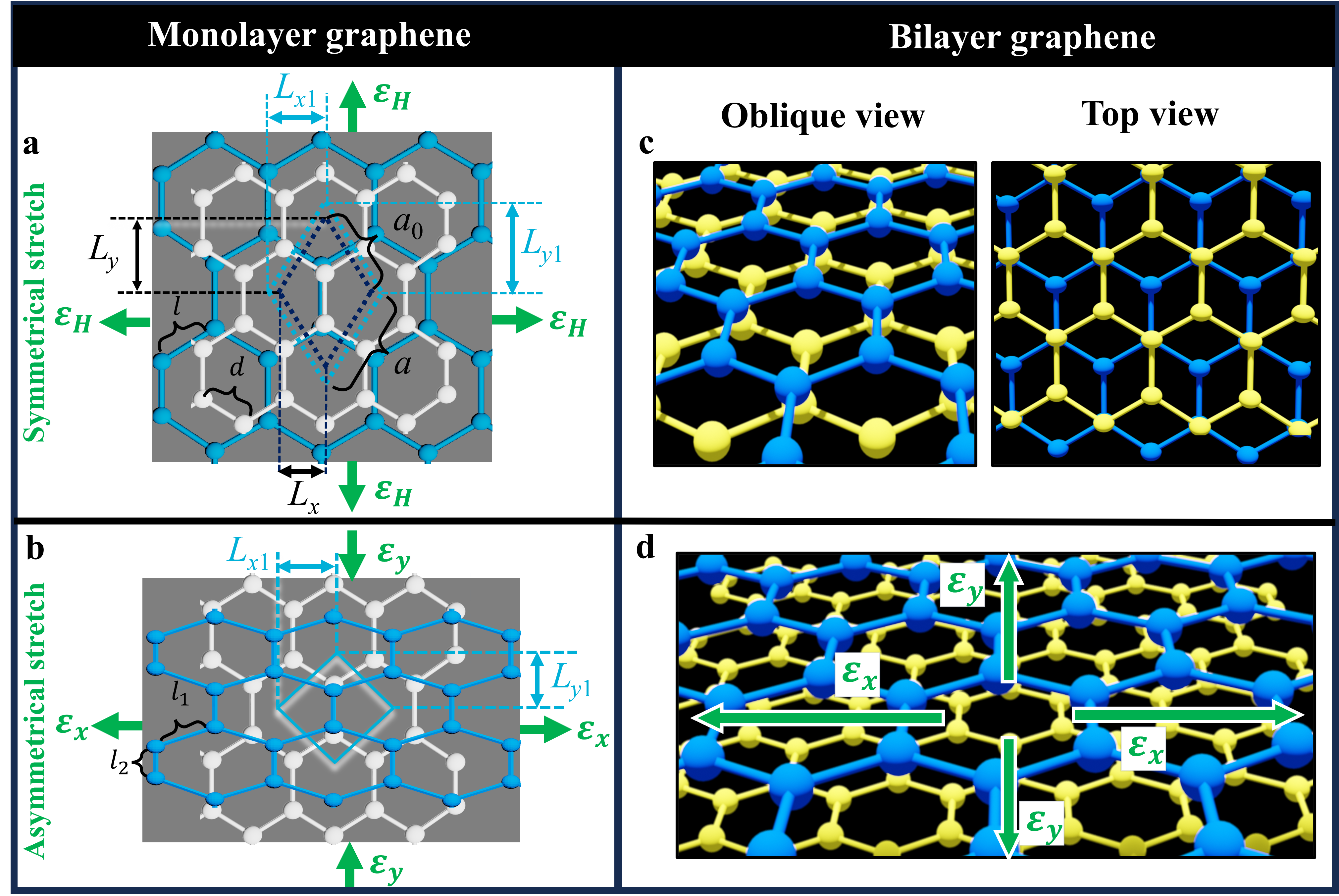}
\caption{\textbf{Relaxation-strain definition.}
Schematic representation of monolayer graphene with \textbf{a} symmetrical deformation  ($\varepsilon_x=\varepsilon_y$) and
\textbf{b} asymmetrical deformation along $x-$ (or $y-$) direction ($\varepsilon_x\neq\varepsilon_y$). Here, the graphene in white (blue) is pristine (deformed).
Diagram of bilayer graphene: \textbf{c} pristine; \textbf{d} deformed. Here $L_x$ ($L_{x_1}$) and $L_y$ ($L_{y_1}$) are the half diagonal lengths of the pristine (deformed) cells, respectively.}
\label{fig-cell}
\end{figure}

(i) symmetrical strain distribution which retains hexagonal symmetry and is defined as $\varepsilon_H=(a-a_0)/a_0$. The terms $a$ and $a_0$ denote the lattice parameters before and after deformation, respectively ( Figure \ref{fig-cell}a); \\

(ii) asymmetrical strain distribution along $x-$ (or $y-$) direction which corresponds to strain parallel to the zigzag (or armchair) edge of graphene ribbons and is defined as $\varepsilon_x=(L_{x1}-L_x)/L_x$  (or $\varepsilon_y=(L_{y1}-L_y)/L_y$) and $\varepsilon_x\neq\varepsilon_y.$  Here $L_x$ ($L_{x 1}$) and $L_y$ ($L_{y 1}$) are the half diagonal lengths of the pristine (deformed) cells (Figure \ref{fig-cell}b). \\

\begin{figure}[!ht]
\centering
\includegraphics[width=\linewidth]{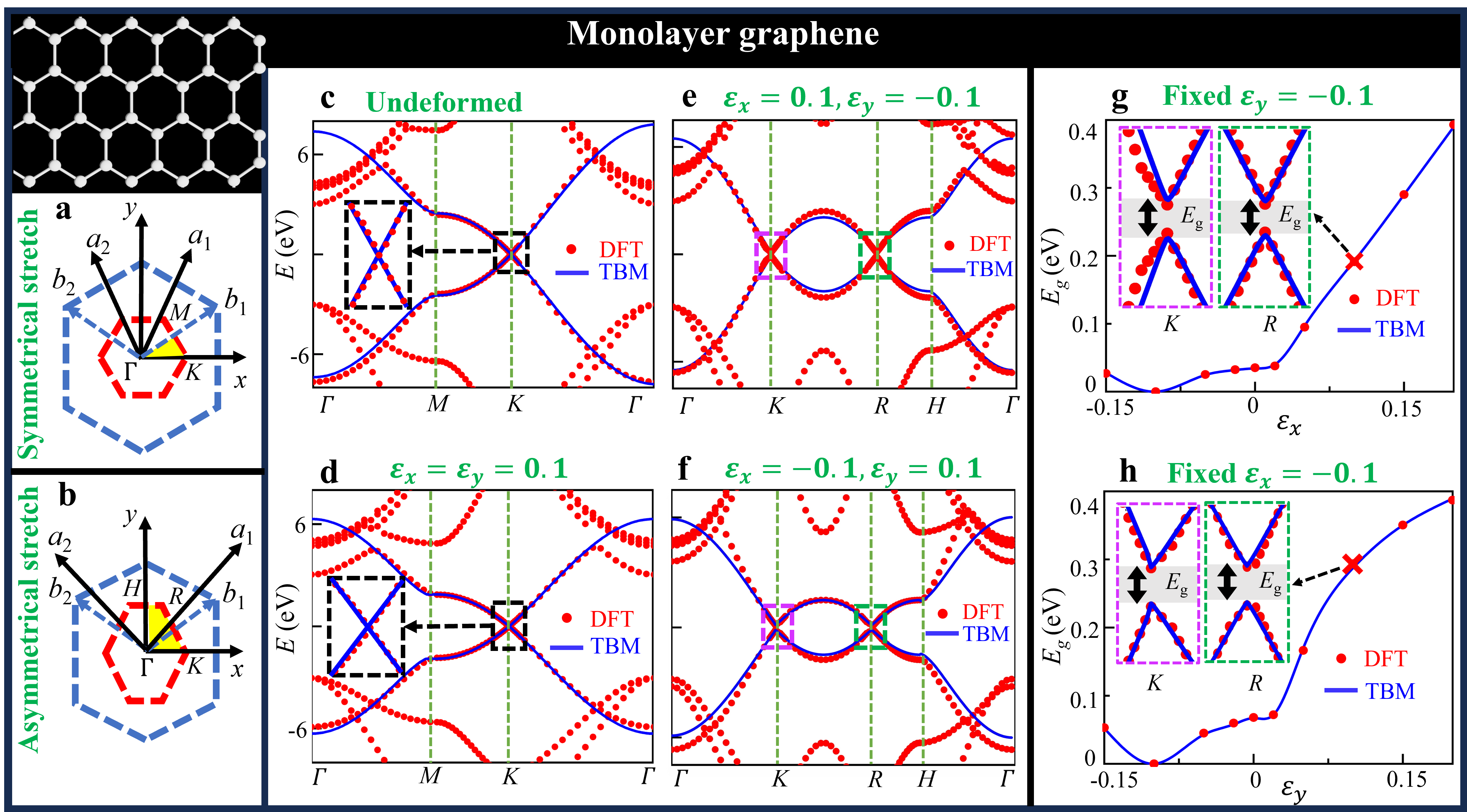}
\caption{\textbf{Semimetal-insulator transition for monolayer graphene.}
Reciprocal lattices (in blue dashed lines), Brillouin zones (in red dashed lines), and irreducible Brillouin zones (in yellow) for monolayer graphenes: \textbf{a} pristine; 
\textbf{b} deformed. The terms $\Gamma$, $K$, $M$, $R$ and $H$ denote the highly symmetrical points.
TBM and DFT of strained monolayer graphene: \textbf{c} pristine; 
\textbf{d} deformed ($\varepsilon_H=0.1$). The insets show zooms around the $K$ point, no bandgaps are induced in both cases.
TBM and DFT of asymmetrical strained monolayer graphene: 
\textbf{e} along $x$ direction ($\varepsilon_x=0.1$, $\varepsilon_y=-0.1$);
\textbf{f} along $y$ direction ($\varepsilon_x=-0.1$,$\varepsilon_y=0.1$). \textbf{g, h} The plot of bandgap as the function of $\varepsilon_x$ and $\varepsilon_y$. The insets show bandgaps generated around the $K$ point and $R$ point, which indicate semimetal-insulator transitions. The cross mark corresponds to strain conditions of the insets.}
\label{fig-monolayer}
\end{figure}

We first establish the TBM for symmetrical strained monolayer graphene only considering on-site and nearest-neighbor hoppings, as shown in Figure \ref{fig-cell}a. The Hamiltonian for monolayer graphene can be expressed as
\begin{equation}\label{GrindEQ__c1}
H=-t \sum_{\mathbf{R}} c_A^{\dagger}(\mathbf{R})\left(c_A(\mathbf{R})+c_B\left(\mathbf{R}-\mathbf{a}_1\right)+c_B\left(\mathbf{R}-\mathbf{a}_2\right)\right)+ h.c.,
\end{equation}
where $c_A^{\dagger}(\mathbf{R})$ and $c_A(\mathbf{R})$ are creation (annihilation) operators for an electron in an atomic-like state of kind A (i.e., three adjacent carbon atoms forms a regular triangle). The terms $a_1$ and $a_2$ are basis vectors for the unit cell, \textbf{R} is the position of the unit cell, and $h.c.$ stands for hermitian conjugate.
We obtain the Hamiltonian for symmetrical strained monolayer graphene
$H_{SMG}(\mathbf{k})$ by
\begin{equation}\label{GrindEQ__c2}
H_{SMG}(\mathbf{k})=\left[\begin{array}{cc}0 & -t f(\mathbf{k}) \\ -t f^*(\mathbf{k}) & 0\end{array}\right],
\end{equation}
where we have $f(\mathbf{k})=\sum_{i=1}^3 e^{i \mathbf{k} \cdot \mathbf{d}_i}$ and $d_1$ = ($a_1$ + $a_2$)($1+\varepsilon_H$) /3, $d_2$ = (-2$a_1$ + $a_2$)($1+\varepsilon_H$) /3, $d_3$ = ($a_1$-2$a_2$)($1+\varepsilon_H$) /3.
After imposing symmetrical strain distributions, the hopping parameters $t$ with the bond length is expressed as $V_{p p \pi}(l)=t_0 e^{-3.37\left(l / d-1\right)}$, where $d$ is the c-c bond length for undeformed graphene \cite{wong2012strain,catarina2019twisted}.

For asymmetrical strained monolayer graphene (Figure \ref{fig-cell}b), the Hamiltonian changes its form to
\begin{equation}\label{GrindEQ__c3}
H_{ASMG}(\mathbf{k})=\left[\begin{array}{cc}0 & -(t_2-t_1)-t_1 f(\mathbf{k}) \\ (-t_2-t_1)-t_1 f^*(\mathbf{k}) & 0\end{array}\right],
\end{equation}
where the new added terms $t_1=V_{p p \pi}(l_1)$ and $t_2=V_{p p \pi}(l_2)$ denote the hopping parameters.

We then move to construct TBM of bilayer graphene considering only a homogeneous interlayer hopping between the nearest neighbors, as shown in Figure \ref{fig-cell}c. The Hamiltonian can be written as the sum of the following terms
\begin{equation}\label{GrindEQ__c4}
H=H_1+H_2+\left\langle 1, \mathbf{R}, A\left|H_{\perp}\right| 2, \mathbf{R}, B\right\rangle \sum_{\mathbf{R}} c_{1, A}^{\dagger}(\mathbf{R}) c_{2, B}(\mathbf{R})+ h.c. ,
\end{equation}
where $H_1$ and $H_2$ are the Hamiltonian for each monolayer graphene, while $H_\perp$ indicates Hamiltonian interlayer coupling in the second quantized formalism. 
The Hamiltonian $H_{BLG}(k)$ of Bernal-stacked bilayer graphene is 
\begin{equation}\label{GrindEQ__c5}
H_{BLG}(\mathbf{k})=\left[\begin{array}{cccc}0 & -t f(\mathbf{k}) & 0 & \left\langle 1, \mathbf{R}, A\left|H_{\perp}\right| 2, \mathbf{R}, B\right\rangle \\ -t f^*(\mathbf{k}) & 0 & 0 & 0 \\ 0 & 0 & 0 & -t f(\mathbf{k}) \\ \left\langle 1, \mathbf{R}, A\left|H_{\perp}\right| 2, \mathbf{R}, B\right\rangle & 0 & -t f^*(\mathbf{k}) & 0\end{array}\right].
\end{equation}
For bilayer graphene systems with bi-axial deformation (shown in Figure \ref{fig-cell}d), we construct a low-energy continuum model that consists of three terms: two single-layer Dirac–Hamiltonian terms that account for the isolated graphene sheets, and a tunneling term that describes hopping between the two layers. 
Considering only the $K$ points of three closest neighbors, we can get Hamiltonian $H_{SBLG}(k)$ for bilayer-strained graphene as
\begin{equation}\label{GrindEQ__c9}
H_{SBLG}(k)=\left[\begin{array}{cccc}H_{MG}^k(\frac{\varepsilon_x}{2}, \frac{\varepsilon_y}{2}) & T_{\mathbf{q}_b} & T_{\mathbf{q}_{t r}} & T_{\mathbf{q}_{t t}} \\ T_{\mathbf{q}_b}^{\dagger} & H_{MG}^{k_b}\left(-\frac{\varepsilon_x}{2},-\frac{\varepsilon_y}{2}) \right) & 0 & 0 \\ T_{\mathbf{q}_{t r}}^{\dagger} & 0 & H_{MG}^{k_{tr}}\left(-\frac{\varepsilon_x}{2},-\frac{\varepsilon_y}{2}\right) & 0 \\ T_{\mathbf{q}_{t l}}^{\dagger} & 0 & 0 & H_{MG}^{k_{tl}}\left(-\frac{\varepsilon_x}{2},-\frac{\varepsilon_x}{2}\right)\end{array}\right].
\end{equation}
Here $H_{MG}$ is the Hamiltonian for monolayer graphene, i.e., $H_{SMG}(k)$ for symmetrical and $H_{ASMG}(k)$ for asymmetrical systemd, $T$ is the tunneling term for interlayer hopping. On basis of the hamiltonian matrix, we further obtain the renormalization of Fermi velocity $v_F^*$:
\begin{equation}\label{GrindEQ__c10}
\frac{v_F^*(\theta)}{v_F}=1-\left(\frac{t_{\perp}(|\mathrm{K}|)}{v_F \hbar|\mathrm{K}| A_{\text {u.c.}}}\right)^2 \frac{1}{\sqrt{(\varepsilon_x^2+\varepsilon_y^2)/2}},
\end{equation}
where $A_{u. c.}$ is unit cell area,  $t_{\perp(K)} = 0.58 \mathrm{~eV} \text {Å}^2$ denotes the interlayer hopping term for Bernal stacked bilayer graphene, $V_F$ is the pristine Fermi velocity, and $\hbar$ is the Plank constant. Following Eq. \ref{GrindEQ__c10}, Fermi velocity will decay to zero under small $\varepsilon_x$ and $\varepsilon_y$, which potentially generates superconductivtiy. Details of the TBM are presented in the supplementary information.

\section*{Results and Discussion}
\subsection*{Semimetal-insulator transition in monolayer graphene}
To verify the accuracy of the established TBM, we conduct simulations based on first principle calculations of density functional theory (DFT). The results by TBM and DFT simulations show perfect agreement with each other, as shown in Figure \ref{fig-monolayer}. In the symmetrical strain conditions ($\epsilon_x= \epsilon_y\neq0$), we observe that the slope of the band structure decreases near the $K$ point, which indicates the decrement of Fermi velocity according to the law $V_F=2 \pi E /(\hbar \cdot k)$. In addition, the bandgap is observed to be zero, because the symmetrical strain field retains the geometry symmetry of hexagonal lattices.
In the asymmetrical strain conditions ($\epsilon_x \neq\epsilon_y\neq0$), the bandgap will open near the $K$ and $R$ high symmetry points, due to the destruction of geometry symmetry in Hexagonal lattices, as observed in Figure \ref{fig-monolayer}e-f. Such a bandgap-opening phenomenon indicates that the monolayer graphene generates semimetal-insulator transitions. Partial enlargement of these band structures are  presented as inserts in Figure \ref{fig-monolayer}g-h which depicts the general relationship between the bandgap and the strain. Results show that the bandgap will open largely if the monolayer graphene is stretched in one direction while compressed in another direction, i.e., inhomogeneous strain condition $\epsilon_x \epsilon_y<0$. It is also found that the bandgap value increases with the increase of applied strain differences. We get a bandgap of $0.39\rm{~eV}$ when strain condition $\epsilon_x=-10\%, \epsilon_y=20\%$ is imposed. This value is much larger than unidirectional stretch or compression obtained in literature \cite{wong2012strain}. In addition, by releasing homogeneous strain in orthogonal directions (compressive strain only or tensile strain only, $\epsilon_x \epsilon_y>0$), we can still obtain bandgap opening, but smaller than the inhomogeneous strain conditions. This can be intuitively interpreted from the fact that inhomogeneous strain conditions will result in a larger destruction of the geometry symmetry in Hexagonal lattices. 

\subsection*{Unconventional superconductivity in bi-layer graphene}
\begin{figure}[!ht]
\centering
\includegraphics[width=\linewidth]{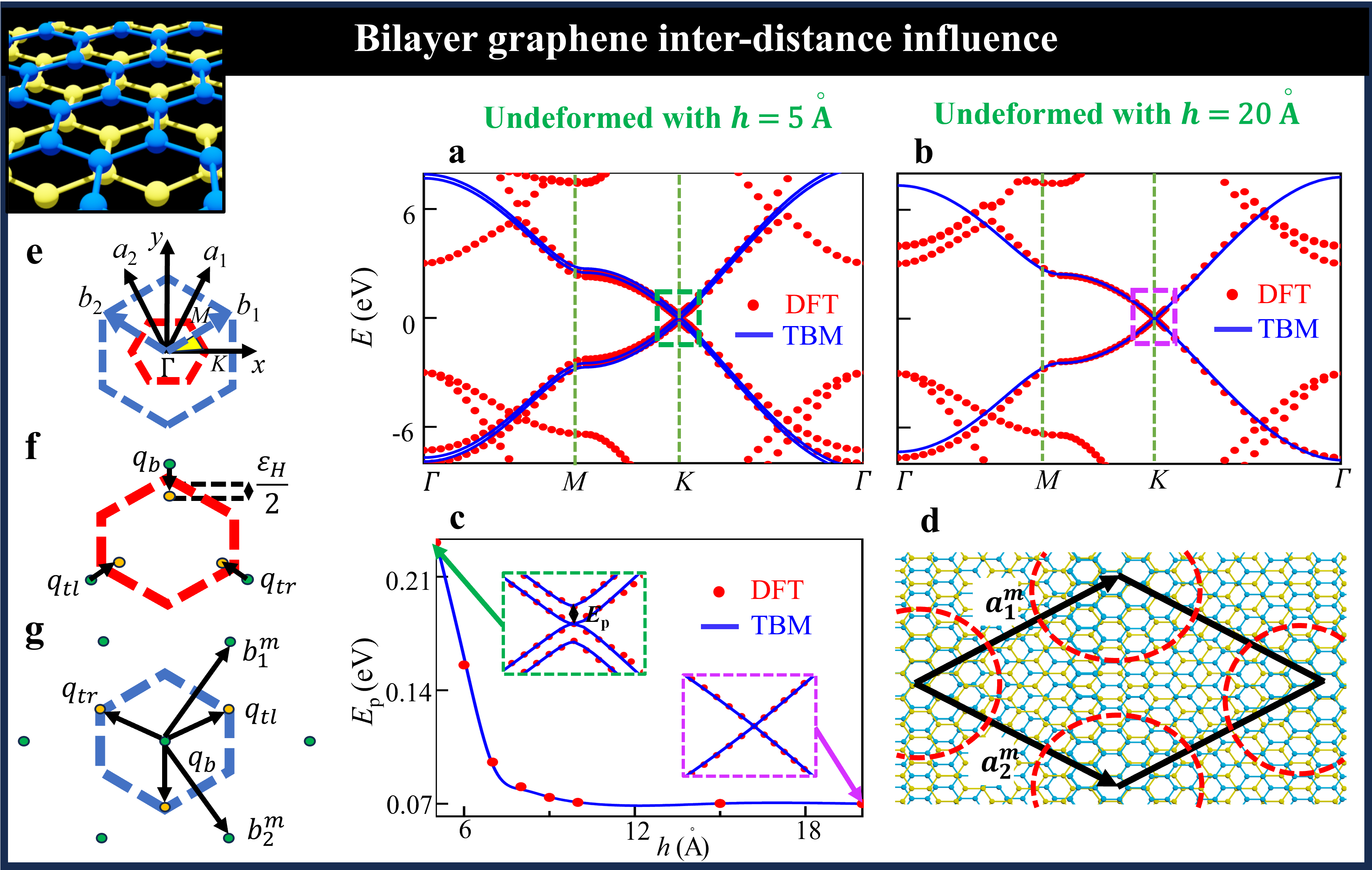}
\caption{\textbf{Interlayer distance effects for bilayer graphene.}
TBM and DFT of Bernal-stacked bilayer graphene with different interlayer distances: 
\textbf{a} $h=5\text {Å}$; 
\textbf{b} $h=20\text {Å}$. The insets show zooms around the $K$ point. The valence band and conduction band are separated from each other, due to the false van der Waals force.
\textbf{c} The separation value of $E_p$ as a function of the interlayer distance $h$.
\textbf{d} Moiré patterns in symmetrical strained  bilayer graphene ($\varepsilon_H=11.3\%$). The red circles denote high-energy AA stacking regions and the black diamond shows potential periodic
computational domain. 
\textbf{e} Reciprocal lattices for bilayer graphene system.
\textbf{f} Momentum-space diagram for the interlayer hopping on a symmetrical strained bilayer graphene. The first Brillouin zone is depicted by red lines for primitive bilayer graphene. The equivalent Dirac points ($K$ and $K'$) are marked by green (orange) dots.
\textbf{g} Three distinct hopping processes in reciprocal space is depicted by  $q_b$, $q_{tr}$ and $q_{tl}$. The blue dashed line marks a moiré unit cell, $b_1^m$ and $b_2^m$ are the basis vectors.}
\label{fig-bilayer}
\end{figure}

We then investigate the band structures of Bernal-stacked bilayer graphene. The influence of interlayer distance is first studied based on TBM and DFT methods. Results in Figure \ref{fig-bilayer}a-b show that the valence band and conduction band will get separated when the interlayer distance is $h=5\text {Å}$. Such phenomenon is induced by the false Van der Waals force, and the result agrees well with literature \cite{catarina2019twisted}. We further reveal the dependence of such separation $E_g$ on the interlayer distance in Figure \ref{fig-bilayer}c. With the increment of interlayer distance, the separation becomes smaller and tends to be negligible when the interlayer distance is larger than 20$\text {Å}$. In subsequent analysis, we consider bilayer graphene with the interlayer distance $h=20\text {Å}$, to eliminate the false wander wales force.
Specifically, we consider deformed bilayer graphene with one layer fixed and another layer stretched or compressed in orthogonal directions ($\varepsilon_H=11.3\%$), as shown in Figure \ref{fig-bilayer}d. The region enclosed by black lines is the unit cell, where $a_1^m$ and $a_2^m$ are the basis vectors, $q_b$, $q_{tr}$ and $q_{tl}$ represent the momentum difference of the $K$ point between the fixed layer and the biaxially stretched layer, as shown in Figure \ref{fig-bilayer}e. We consider the $K$ points of the nearest three neighbours in the fixed layer  (Figure \ref{fig-bilayer}f), their momentum differences to the origin exactly meet the momentum conservation law.
Besides, the $K$ points are staggered due to biaxial stretch, they constitute a new set of honeycomb lattices, thus satisfying the requirements by equation  (\ref{GrindEQ__c9}).

\begin{figure}[!ht]
\centering
\includegraphics[width=\linewidth]{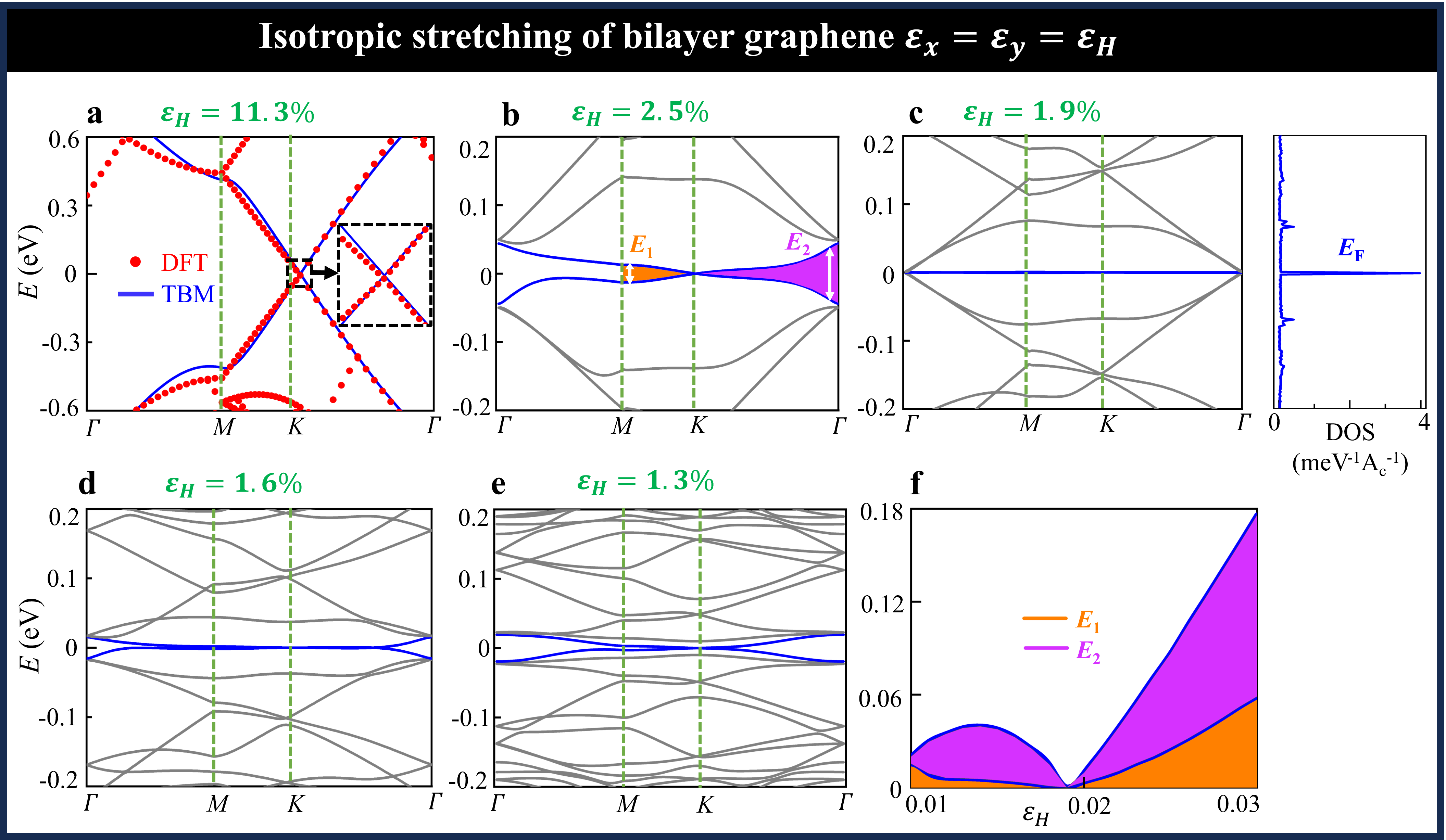}
\caption{\textbf{Superconductity in magic-strain bilayer graphene}.
TBM and DFT of the symmetrical strained bilayer graphene with different stretching strains: 
\textbf{a} $\varepsilon_H=11.3\%$; 
\textbf{b} $\varepsilon_H=2.5\%$. 
The insets show zooms around the $K$ point. No bandgap is observed, and the Fermi velocity $V_F=2 \pi E /(\hbar k)$ decreases with decreasing stretching strains, as the curve slopes around the $K$ point decrease in the range $\varepsilon_H>1.9\%$. \textbf{c} For strain condition $\varepsilon_H=1.9\%$, flat band is observed in the left panel, and a peak value appears at the Fermi level in the right panel, which demonstrates potential superconductivity.
\textbf{d} $\varepsilon_H=1.6\%$; 
\textbf{e} $\varepsilon_H=1.3\%$.
\textbf{f} Band gap at $\Gamma$ and $M$ points is plotted as the function of strain $\varepsilon_H$.
The bandgap first gradually decays to zero at $\varepsilon_H=1.9\%$, then increases beyond this critical value.}
\label{fig-bilayer strain}
\end{figure}

We compare the TBM and DFT results for bilayer graphene with bi-axial symmetrical deformation ($\epsilon_x= \epsilon_y\neq0$). As compression can easily induce wrinkling of graphene sheets in practical senses \cite{androulidakis2017wrinkling} and affect the electrical properties, here we only emphasize stretch conditions. Analysis on compression conditions are in the supplementary information. As shown in Figure \ref{fig-bilayer strain}a, theoretical predictions (by TBM) are in good agreement with simulation results (by DFT), verifying the effectiveness of our TBM. Based on the established TBM, we first investigate the band structures of bilayer graphene with different stretch conditions. It is found that the curve slope near the $K$ point decreases gradually with reduced tensile strain, which indicates a growing lower Fermi velocity. In Figure \ref{fig-bilayer strain}c, we show the band structure and density of states near the charge neutrality point calculated for $\varepsilon_H=1.9\%$. 
A flat band is observed for the band structure and a peak value appears for the density of states at the Fermi level, which indicates that the Fermi velocity of the electron is zero, {i.e., a magic strain in analogy with magic angle is obtained. In this {magic-strain case, it is difficult for the electron to hop from the conduction band to the valence band. We adopt the McMillan formula \cite{mcmillan1968transition} to obtain the Bardeen–Cooper–Schrieffer (BCS) superconductivity critical temperature as $T_c=\frac{\hbar \omega_D}{1.45 k_B} \exp \left(-\frac{1.04 (1+\lambda)}{\lambda-\mu_c^*(1+0.62 \lambda)}\right)$, here, $\lambda$ is a strong BCS coupling strength and is slightly larger than 1, $\hbar \omega_D$ is the Debye frequency, and $\mu_c^*$ is the reduced Coulomb coupling strength. If we use $\lambda=1.3$, the BCS superconductivity critical temperature $T_c$ will be  $0.53\rm{~K}$. For any operation temperatures below $T_c$, superconductivity will be generated. Note that we did not fulfill the tough task of exactly calculating $\lambda$ and $T_c$ for our system, but made estimations of  $T_c$ on typical $\lambda$ instead. We aim to demonstrate that the high density of states at the Fermi level will induce a strong phonon-electron coupling which can cause superconductivity. Such estimations are enough for proof-of-concept demonstration. See more details in the supplementary information. As a quantitative illustration, we present the relationship between the bandgap at $\Gamma$ and $M$ points in Figure \ref{fig-bilayer strain}f. It is observed that only under the magic strain $\varepsilon_H=1.9\%$, the D-value (refers to the difference between $E_1$ and $E_2$) approaches zero, as verified by Figure \ref{fig-bilayer strain}c.

We further investigate the bilayer graphene system with bi-axial asymmetrical strains ($\epsilon_x \neq \epsilon_y$). It is found that the bandgaps are open only under asymmetrical strain conditions and are closed if symmetrical strains are imposed, as shown in  Figure \ref{fig-bilayer asstrain}a-e. The density of states is shown in the right panel of Figure \ref{fig-bilayer asstrain}c, where neither peak value nor bandgap are observed. In Figure \ref{fig-bilayer asstrain}f we show that the bandgap value increases with increased D-value of tensile strain in $x$ and $y$ direction, as a result of increased destruction of geometry symmetry. In the case where $\varepsilon_x =-0.15$ and $\varepsilon_y=0.15$, the value of bandgap is observed to be $0.0272\rm{~eV}$ which is much smaller than its strained monolayer graphene counterpart. Such findings indicate that monolayer graphene is much easier to generate semimetal-insulator transition than bilayer graphene when relaxation strains are imposed. In addition, the bandgap value exhibits different dependency behavior on $\varepsilon_x$ and $\varepsilon_y$, which is induced by the chiral properties of graphene.

\begin{figure}[!ht]
\centering
\includegraphics[width=\linewidth]{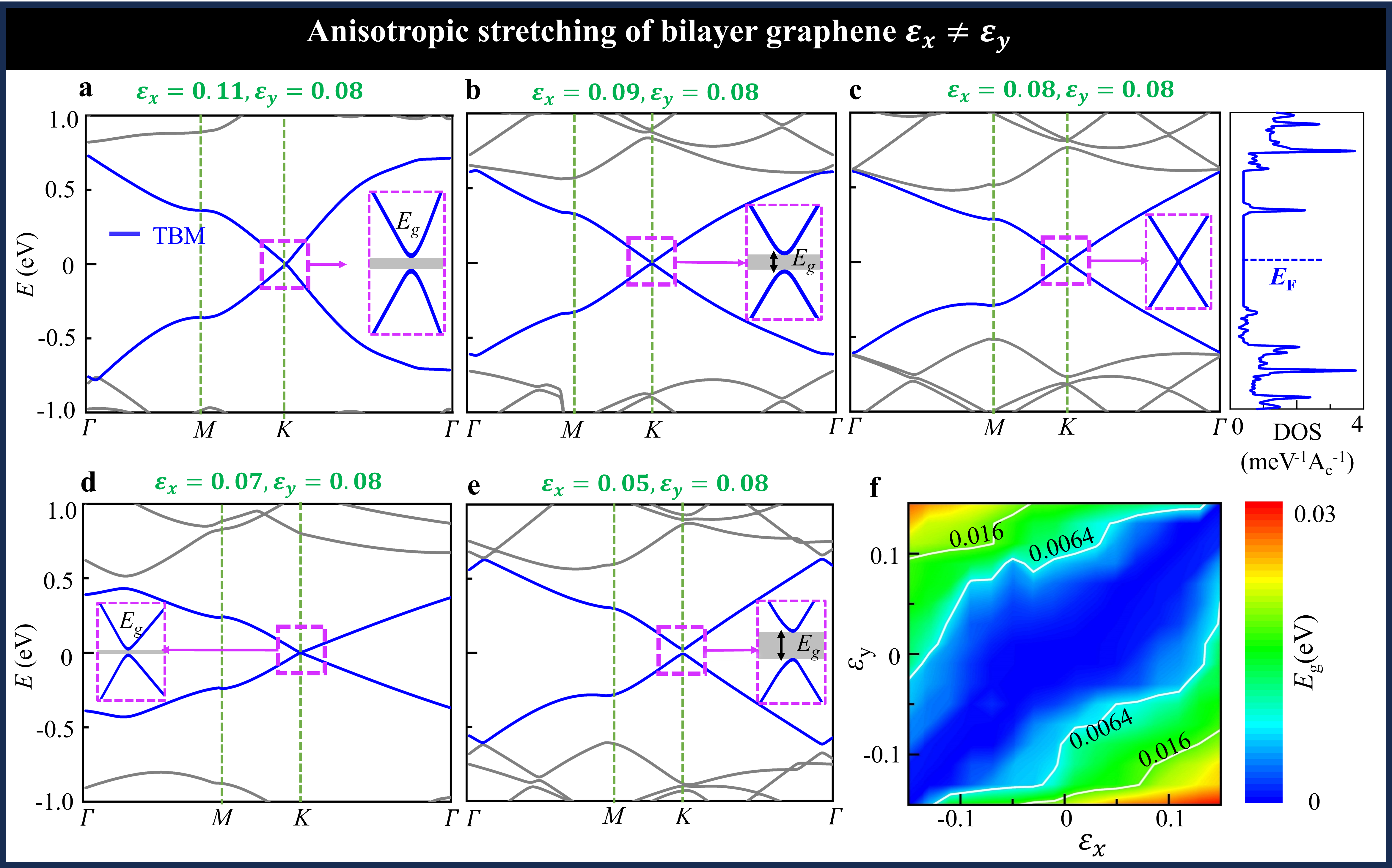}
\caption{\textbf{Band structure for anisotropic strained bilayer graphene.}
TBM of the asymmetrical strained bilayer graphene with different strain conditions:  
\textbf{a} $\varepsilon_x=0.11$; 
\textbf{b} $\varepsilon_x=0.09$; 
\textbf{c} $\varepsilon_x=0.08$;
\textbf{d} $\varepsilon_x=0.07$;
\textbf{e} $\varepsilon_x=0.05$. For each system $\varepsilon_y=0.08$ is imposed.
The insets show zooms around the $K$ point. Bandgaps, which indicate semimetal-insulator transitions, are generated except for the symmetrical condition.  \textbf{f} Bandgap at $K$ point is plot as functions of $\varepsilon_x$ and $\varepsilon_y$. The bandgap will increase with increasing D-value of $\varepsilon_x$ and $\varepsilon_y$.}
\label{fig-bilayer asstrain}
\end{figure}

\section*{Conclusion}
\noindent Summarizing, we have shown that relaxation-strained graphene has the potential to be a superconductor or insulator. Firstly, asymmetrical strain distribution will result in the bandgap opening of monolayer graphene, which indicates that a semimetal-insulator transition is generated. If we impose different types of strain on the monolayer graphene (compressive strain in one direction and tensile strain in another direction, $\epsilon_x \epsilon_y<0$), the bandgap will open largely due to severe destruction of the geometry symmetry in hexagonal lattices. By contrast, if the same types of strain are applied (compression or stretch in both directions, $\epsilon_x \epsilon_y>0$ ), the bandgap of monolayer graphene is small. In extreme conditions, if the stretch or compression rates in two directions are identical ($\epsilon_x =\epsilon_y$), the bandgap will vanish, also the curve slope near the $K$ point will be reduced relative to the pristine graphene, which indicates that stretch or compression will reduce the Fermi velocity. Following these findings, we compression the monolayer graphene by $10\%$ in one direction and stretch it by different rates in another direction. It is found that the bandgap value increases with the increase of strain differences. Specifically, the bandgap can be as large as $0.39\rm{~eV}$ on condition of $\epsilon_x=-10\%, \epsilon_y=20\%$, which is much larger than unidirectional stretch or compression ever reported.
Secondly, a small interlayer distance will induce separation of the conduction band and valence band due to false van der Walls forces, and such separation phenomenon can be eliminated if the interlayer distance is larger than $10 \text{Å}$. 
Lastly, under the condition that one graphene layer is fixed while another layer is bi-axially stretched (or compressed), the Fermi velocity will decrease with decreasing tensile strains. When the symmetrical strain is at the magic-strain $1.9\%$, a flat band is generated which indicates that the bilayer graphene turns out to be a superconductor below the critical temperature. By contrast, bi-axially asymmetrical stretched (or compressed) conditions will generate bandgap opening which indicates semimetal-insulator transitions. Generally, we pave a new avenue to achieve graphene superconducting or insulating states by tailoring bi-axial strains. Compared with widely-used twisted systems, the relaxation strain is easier to implement in practice and adds more flexibility to obtain exotic electronic properties by strain engineering. 

\section*{Theory and calculation}
\subsection*{Theoretical model}
\noindent Considering the situation that one graphene layer is fixed and another layer is stretched in orthogonal directions, the Hamiltonian will consist of two single-layer Dirac–Hamiltonian terms and a tunneling term. In this section, we present the process to simplify the tunneling term. 
The matrix element for the tunneling term based on the continuum model is
\begin{equation}\label{GrindEQ__c11}
T_{\mathbf{k},\mathbf{k}^{\prime}}^{\alpha,\beta}=\left\langle\Psi_{\mathbf{k},\alpha}\left|H_{\perp}\right| \Psi_{\mathbf{k}^{\prime},\beta}^{\varepsilon}\right\rangle.
\end{equation}
Here, the tunneling Hamiltonian $H_{\perp}$ describes a process during which an electron with momentum $\mathbf{k}^{\prime}=M \mathbf{k}$ in the fixed layer hops to the momentum state ${\bf k}$ in the stretched layer. The left and right vectors are Bloch wave functions

\begin{equation}
\begin{array}{cc}
    \left|\psi_{\mathbf{k}, \alpha}\right\rangle &=\frac{1}{\sqrt{N_1 N_2}} \sum_{n_1, n_2} e^{i \mathbf{k} \cdot\left(\mathbf{R}_{n_1, n_2}+\mathbf{\delta}_\alpha\right)}\left|\mathbf{R}_{n_1, n_2}+\mathbf{\delta}_\alpha, \alpha\right\rangle, \\ 
    
\left|\psi_{\mathbf{k}^{\prime}, \beta}^{\varepsilon}\right\rangle
&=\frac{1}{\sqrt{N_1 N_2}} \sum_{n_1^{\prime}, n_2^{\prime}} e^{i \mathbf{k}^{\prime} \cdot\left(\mathbf{R}_{n_1^{\prime}, n_2^{\prime}}^{\varepsilon}+\mathbf{\delta}_\beta^{\varepsilon}\right)}\left|\mathbf{R}_{n_1^{\prime}, n_2^{\prime}}^{\varepsilon}+\mathbf{\delta}_\beta^{\varepsilon}, \beta\right\rangle. 
\label{GrindEQ__c12}
\end{array}
\end{equation}
where the vectors in the deformed layer have all taken into account the tensile strain, and they are set as $\alpha=A, \delta_\alpha=0$ and $\alpha=B,  
\delta_\alpha=\delta$. Substituting equation (\ref{GrindEQ__c12}) into equation (\ref{GrindEQ__c11}), we can obtain 

\begin{equation*}
\label{GrindEQ__c13}
\begin{array}{cc}
 T_{K+\mathbf{q}_1, K^\varepsilon+\mathbf{q}_2^\varepsilon}^{\alpha, \beta}&= \frac{1}{N_1 N_2} \sum_{n_1, n_2} \sum_{n_1^{\prime}, n_2^{\prime}} e^{-i\left(K+\mathbf{q}_1\right) \cdot\left(\mathbf{R}_{n_1, n_2}+\mathbf{\delta}_\alpha\right)} e^{i\left(K^\varepsilon+\mathbf{q}_2^\varepsilon\right) \cdot\left(\mathbf{R}_{n_1^{\prime}, n_2^{\prime}}^{\varepsilon}+\mathbf{\delta}_\beta^{\varepsilon}\right)} \\
 \vspace{0.5cm}
     & \times \left\langle\mathbf{R}_{n_1, n_2}+\mathbf{\delta}_\alpha, \alpha\left|H_{\perp}\right| \mathbf{R}_{n_1^{\prime}, n_2^{\prime}}^{\varepsilon}+\mathbf{\delta}_\beta^{\varepsilon}, \beta\right\rangle .
\end{array}
\end{equation*}

We define the last term as a transition matrix element
\begin{equation}\label{GrindEQ__c14}
\left\langle\mathbf{R}_{n_1, n_2}+\mathbf{\delta}_\alpha, \alpha\left|H_{\perp}\right| \mathbf{R}_{n_1^{\prime}, n_2^{\prime}}^{\varepsilon}+\mathbf{\delta}_\beta^{\varepsilon}, \beta\right\rangle=t_{\perp}\left(\mathbf{R}_{n_1, n_2}+\mathbf{\delta}_\alpha-\mathbf{R}_{n_1^{\prime}, n_2^{\prime}}^{\varepsilon}-\mathbf{\delta}_\beta^{\varepsilon}\right),
\end{equation} 
and use Fourier transform for simplification 
\begin{equation}
\label{GrindEQ__c15}
\begin{array}{cc}
 T_{K+\mathbf{q}_1, K^\varepsilon+\mathbf{q}_2^\varepsilon}^{\alpha, \beta}=    & \frac{1}{\left(N_1 N_2\right)^2} \sum_{n_1, n_2} \sum_{n_1^{\prime}, n_2^{\prime}} \sum_{\mathbf{k}} e^{i\left[\mathbf{k}-\left(K+\mathbf{q}_1\right)\right] \cdot \mathbf{R}_{n_1, n_2}} \times e^{i\left[\left(K^\varepsilon+\mathbf{q}_2^\varepsilon\right)-\mathbf{k}\right] \cdot \mathbf{R}_{n_1^{\prime}, n_2^{\prime}}^\varepsilon}  \\
     & \times e^{i\left[\mathbf{k}-\left(K+\mathbf{q}_1\right)\right] \cdot \mathbf{\delta}_\alpha+\mathbf{\tau}\times e^{i\left[\left(K^\varepsilon+\mathbf{q}_2^\varepsilon\right)-\mathbf{k}\right]\cdot\left(\mathbf{\delta}_\beta^\varepsilon-\mathbf{\delta}+\mathbf{\tau}\right)} \frac{t_{\perp}(\mathbf{k})}{A_{\rm u . c .}}} .
\end{array}
\end{equation}

We then define reciprocal lattice vectors to simplify equation (\ref{GrindEQ__c15}), and transform its form from the real space to the reciprocal space:
\begin{equation}\label{GrindEQ__c16}
T_{K+\mathbf{q}_1, K^\varepsilon+\mathbf{q}_2^\varepsilon}^{\alpha, \beta}=\sum_{k,l,m,n} \frac{t_{\perp}\left(K+\mathbf{q}_1+\mathbf{G}_{k, l}\right)}{A_{\rm u . c .}} e^{i\left[\mathbf{G}_{k, l} \cdot \mathbf{\delta}_\alpha-\mathbf{G}_{m, n} \cdot\left(\mathbf{\delta}_\beta^\varepsilon-\mathbf{\delta}\right)-\mathbf{G}_{m, n}^\varepsilon \cdot \mathbf{\tau}\right]} \delta_{K+\mathbf{q}_1+\mathbf{G}_{k, l}, K^\varepsilon+\mathbf{q}_2^\varepsilon+\mathbf{G}_{m, n}^\varepsilon}.
\end{equation}
Here, G is summed over reciprocal lattice vectors. The main contribution sum in the formula $T_{K+q_1, K^\varepsilon+q_2^\varepsilon}^{\alpha, \beta}$ originates from $G_{m,n}$, $b_2^{\varepsilon}$ and $-b_1^{\varepsilon}$, hence $K+G_{m,n}^{\varepsilon}$ correspond to three K points. In this manner, $q_1$ and $q_2^{\varepsilon}$, which are close to $K$ and $K^{\varepsilon}$, can satisfy the momentum conservation law. Substituting the value $G_{m,n}$ into above equations, we then obtain 
\begin{equation}\label{GrindEQ__c17}
\begin{array}{cc}
T_{K+\mathbf{q}_1, K^\varepsilon+\mathbf{q}_2^\varepsilon}^{\alpha, \beta}= &\frac{t_{\perp}(K)}{A_{u . c .}}[\delta_{K+\mathbf{q}_1, K^\varepsilon+\mathbf{q}_2^\varepsilon}+e^{i\left[\mathbf{b}_2 .\left(\mathbf{\delta}_\alpha-\mathbf{\delta}_\beta^\varepsilon+\mathbf{\delta}\right)-\mathbf{b}_2^\varepsilon. \mathbf{\tau}\right]} \delta_{K+\mathbf{q}_1+\mathbf{b}_2, K^\varepsilon+\mathbf{q}_2^\varepsilon+\mathbf{b}_2^\varepsilon} \\
     & +e^{-i\left[\mathbf{b}_1 \cdot\left(\mathbf{\delta}_\alpha-\mathbf{\delta}_\beta^\varepsilon+\mathbf{\delta}\right)-\mathbf{b}_1^\varepsilon\cdot \mathbf{\tau}\right]} \delta_{K+\mathbf{q}_1-\mathbf{b}_1, K^\varepsilon+\mathbf{q}_2^\varepsilon-\mathbf{b}_1^\varepsilon}].
\end{array}
\end{equation}
Here, all the four possible degrees of freedom for the sublattice are $\{\alpha, \beta\}=\{A, B\}$, $\delta_A=0, \delta_B=\delta$. Then we can write the transition matrix in a two-order form
\begin{equation}\label{GrindEQ__c18}
T=\left[\begin{array}{ll}T^{A, A} & T^{A, B} \\ T^{B, A} & T^{B, B}\end{array}\right].
\centering
\end{equation}
Thus we can obtain the simplified tunneling term that describes interlayer hopping as
\begin{equation}\label{GrindEQ__c19}
\begin{array}{cc}
T_{K+\mathbf{q}_1, K^\varepsilon+\mathbf{q}_2^\varepsilon}^{\alpha,\beta}=T_{\mathbf{q}_b} \delta_{\mathbf{q}_2^\varepsilon-\mathbf{q}_1, \mathbf{q}_b}+T_{\mathbf{q}_{t r}} \delta_{\mathbf{q}_2^\varepsilon-\mathbf{q}_1, \mathbf{q}_{t r}}+T_{\mathbf{q}_{t l}} \delta_{\mathbf{q}_2^\varepsilon-\mathbf{q}_1, \mathbf{q}_{tl}},
\end{array}
\end{equation}
where $\delta$ is a vector connecting the two atoms in the unit cell, $\alpha$ and $\beta$ are the sublattice numbers for the fixed layer and stretched layer, respectively. The transition matrices are given by
$$T_{\mathbf{q}_b}=\frac{t_{\perp}(K)}{A_{\text {u.c. }}} \left[\begin{array}{ll}1 & 1 \\ 1 & 1\end{array}\right],$$ 

$$ T_{\mathbf{q}_{t r}}=\frac{t_{\perp}(K)}{A_{u . c .}}e^{-b_2^m \cdot \tau}\left[\begin{array}{cc}e^{-i \theta} & 1 \\ e^{i \theta} & e^{-i \theta}\end{array}\right], $$

$$ T_{\mathbf{q}_{t l}}=\frac{t_{\perp}(K)}{A_{\text {u.c. }}} e^{-b_1^m \cdot \tau}\left[\begin{array}{cc}e^{i \theta} & 1 \\ e^{-i \theta} & e^{i \theta}\end{array}\right],$$
where $\tau$ is a translation vector that is almost zero for a small stretch factor, $b_1^m$ and $b_2^m$ are basis vectors for reciprocal lattices shown in Figure \ref{fig-bilayer}e.
 Details on the establishing process of the TBM for different graphene systems are provided in the Supplementary Information file.

\subsection*{Density Function Theory Calculation}
\noindent All DFT calculations are conducted using the Vienna Ab initio Simulation package (VASP). The generalized gradient approximation (GGA) and Perdew-Burke Ernzerhof (PBE) function are employed for the exchange-correlation functions. Additionally, the projected augmented wave (PAW) method is utilized to describe the electron interactions. Van der Waals interactions are accounted for using the DFT-D2 method. The truncation energy of plane waves is set to be 550 $\rm{eV}$. Structural optimization is considered complete when the force on each atom is less than 0.01 $\mathrm{~eV}/\text {Å}$. During the process of structural relaxation, a $5\times5\times1$ $K$ point mesh, based on the Monkhorst-Pack scheme, is employed for geometry optimization. Similarly, a $15\times15\times1$ $K$-point mesh is used for electronic structure calculations. Different $K$-point paths are selected based on the specific graphene models under investigation.\\

\section*{Data availability}
All the data supporting the conclusions of this study are included in the article and the Supplementary Information file. Data for the figures can be found in the file of Source Data. Source data are provided in this paper.\\


\section*{CRediT authorship contribution statement}
{\bf Qingxiang Ji}: Conceptualization, Formal analysis, Writing original draft. {\bf Bohan Li}: Formal analysis, Methodology, Software. {\bf Johan Christensen}: Methodology, Data curation, Investigation, Review and editing. {\bf Changguo Wang}: Supervision, Validation, Project administration. {\bf Muamer Kadic}: Conceptualization, Supervision, Validation, Review and editing. \\

\section*{Declaration of competing interest}
The authors declare that they have no known competing financial interests or personal relationships that could have appeared to influence the work reported in this paper.\\

\section*{Acknowledgements}
This work was supported by the National Natural Science Foundation of China [grant numbers 12302169; 12172102], the French Investissements d'Avenir program, in part by the ANR PNanoBot [ANR-21-CE33-0015] and ANR OPTOBOTS project [ANR-21-CE33-0003]. Computations have been performed on the supercomputer facilities of the Mésocentre de calcul de Franche-Comté. \\

\end{document}